\documentclass[%
 reprint,
 amsmath,amssymb,
 aps,
]{revtex4-2}

\newtheorem{corollary}{Corollary}

\newtheorem{proposition}{Proposition}

\usepackage{subfig}
\usepackage{todonotes}
\usepackage{comment}

\usepackage{physics}
\newcommand{\vectorize}[1]{ {\operatorname{vec}\left(#1\right)}}
\newcommand{\swapgate}[1]{ {\operatorname{SWAP}_{#1}}}

\usepackage{graphicx}% Include figure files
\usepackage{dcolumn}% Align table columns on decimal point
\usepackage{bm}% bold math

\begin{document}

\preprint{APS/123-QED}

\title{Distributed quantum logic algorithm}% Force line breaks with \\

\author{Boris Arseniev}
\affiliation{Skolkovo Institute of Science and
Technology, Moscow, Russian Federation}%Lines break 

\date{\today}% It is always \today, today,

\begin{abstract}

Parallel computation enables multiple processors to execute different parts of a task simultaneously, improving processing speed and efficiency.
In quantum computing, parallel gate implementation involves executing gates independently in different registers, directly impacting the circuit depth, the number of sequential quantum gate operations, and thus the algorithm execution time. 
This work examines a method for reducing circuit depth by introducing auxiliary qubits to enable parallel gate execution, potentially enhancing the performance of quantum simulations on near-term quantum devices. 
We show that any circuit on $n$ qubits with depth $O\left(M n^2\right)$, where $M = M(n)$ is some function of $n$, can be transformed into a circuit with depth $O\left(\log_2(M) n^2\right)$ operating on $O\left(M n\right)$ qubits.
This technique may be particularly useful in noisy environments, where recent findings indicate that only the final $O\left(\log n\right)$ layers influence the expectation value of observables. It may also optimize Trotterization by exponentially reducing the number of Trotter steps.
Additionally, the method may offer advantages for distributed quantum computing, and the intuition of treating quantum states as gates and operators as vectors used in this work may have broader applications in quantum computation.
\end{abstract}

\maketitle

%\tableofcontents

\section{\label{sec:intro} Introduction}

Under the assumption that certain quantum gates can be executed in parallel \cite{levine2019parallel, evered2023high}, the depth of a quantum circuit - defined as the number of sequential gate layers required for execution - emerges as a critical parameter in quantum computing. Circuit depth plays a fundamental role in determining the efficiency and feasibility of quantum algorithms. As the circuit depth increases, the execution time increases, leading to an increased risk of decoherence and computational errors. This issue is particularly significant in current quantum hardware, which is constrained by short coherence times that limit the reliable maintenance of quantum states. During algorithm execution, qubits must remain isolated from external interactions to prevent unintended disturbances that could compromise the integrity of computational outcomes \cite{zurek1991decoherence, unruh1995maintaining, palma1996quantum, aharonov1997fault, temme2017error, saki2019study}. These challenges underscore the importance of circuit depth reduction as a key objective in quantum circuit design and optimization \cite{abdessaied2013reducing, amy2013meet, jiang2020optimal, de2021reducing, gyongyosi2020circuit}.  

This work introduces a universal methodology for reducing circuit depth by leveraging additional qubits to enable the parallelization of quantum operations. By defining two specialized gates, $G$ and $V$, we demonstrate how a sequence of $M$ quantum operators can be executed in parallel, resulting in a circuit depth of $O(\log(M))$. This approach becomes especially relevant in noisy quantum environments, as recent findings suggest that only the final $O(\log(n))$ layers of deep circuits significantly affect the expectation values of observables \cite{mele2024noise}. 

The proposed algorithm can also be interpreted within a divide-and-conquer framework \cite{jiang2020optimal}, where segments of the computational task are parallelized while others are executed sequentially. Moreover, it exhibits conceptual similarities to methodologies employed in quantum machine learning, particularly tensor networks, which employ analogous circuit geometries to optimize computational performance \cite{huggins2019towards}. 

This parallelization technique offers substantial promise for near-term quantum (NISQ) devices, where mitigating the impact of short decoherence times remains a critical challenge. By reducing circuit depth and, consequently, execution time, this methodology enables the development of longer and more expressive circuits, making it particularly well-suited for addressing complex computational tasks in the NISQ era \cite{wu2021expressivity}.

A notable application of the proposed algorithm lies in Hamiltonian simulation, where Trotterization protocols often lead to significant increases in circuit depth \cite{berry2007efficient}. Moreover, the ability to perform quantum gate operations in parallel introduces transformative possibilities for distributed quantum computing, allowing for efficient collaboration among multiple quantum processors \cite{caleffi2024distributed}.  

Beyond practical applications, this methodology presents an alternative perspective on quantum circuits, where quantum states are treated as gates and operators as vectors, fostering new perspectives on circuit design and optimization.

The remainder of this paper is organized as follows: Section \ref{sec:build} introduces the core components of the algorithm, including the $G$ and $V$ gates. Section \ref{subsec:dql_alg} details the algorithm, followed by an analysis of operational complexity in Section \ref{subsec:implement}.

\section{Building blocks of the algorithm} \label{sec:build}

The central concept utilized in this work is the application of vectorized operators. We represent the vectorization of a matrix $M$ as $\vectorize{M}$ (sometimes in literature it is also denoted as $|M^T \rangle\rangle$ \cite{braccia2024computing}), which refers to the column vector formed by stacking the columns of matrix $M$ sequentially on top of each other \cite{Phan_2009}. For example

\begin{equation*}\label{eq:example_vec}
\vectorize{
\begin{bmatrix}
    M_{11} & M_{12} \\
    M_{21} & M_{22} \\
\end{bmatrix}
}
= 
\begin{pmatrix}
    M_{11} \\
    M_{21} \\
    M_{12} \\
    M_{22} \\ 
\end{pmatrix}.
\end{equation*}

In this work, we utilize square matrices of dimension $N \times N$, where $N = 2^n$, unless stated otherwise. This implies that each operator acts on $n$ qubits. We also introduce the term \textbf{register} to denote specific subsets of qubits within a tensor product space; generally, each register in this context contains $n$ qubits. For instance, $A \otimes B$ indicates that operator $A$ acts on the first register (the first $n$ qubits), while operator $B$ acts on the second register (the last $n$ qubits).

\begin{figure}%
    \centering
    \subfloat[\centering Sequential circuit]{{\includegraphics[width=0.35\linewidth]{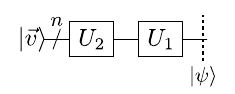} }}%
    \qquad
    \subfloat[\centering Parallel circuit]{{\includegraphics[width=0.45\linewidth]{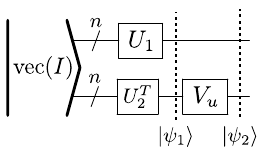} }}%
    \caption{Figure (a) is circuit of 2 gates implemented sequentially, while (b) is the same circuit but implemented in parallel.}%
    \label{fig:example_parseq_circ}
\end{figure}

Before going into the details of the algorithm, we provide a preliminary example, illustrated in Figure \ref{fig:example_parseq_circ}, which presents two circuits for comparison. Circuit (a) consists of two gates applied sequentially, producing the resulting state
\begin{equation*}
    \ket{\psi} = U_1 U_2 \ket{\vec{v}}.
\end{equation*}
In contrast, Figure (b) shows a circuit with two gates applied in parallel, where
\begin{equation*}
    \ket{\psi_1} = \left(U_1 \otimes U_2^T \right) \ket{\vectorize{I}} = \frac{1}{\sqrt{N}} \ket{\vectorize{(U_1 U_2)^T}},
\end{equation*}
and the gate $V_u$, which will be defined in a later section, yields the following result upon application:
\begin{equation*}
\begin{split}
    \ket{\psi_2} &= \frac{1}{\sqrt{N}} \left(I \otimes V_u \right) \ket{\vectorize{(U_1 U_2)^T}} \\
    &= \frac{1}{\sqrt{N}} U_1 U_2 \ket{\vec{v}} \otimes \ket{0} + \sum_{j=1}^{2^n} \ket{\vec{r}} \otimes \ket{j},
\end{split}
\end{equation*}
where $\ket{\vec{r}}$ represents a residual state. Note that a measurement yielding $\ket{0}$ in the second register reproduces the same outcome as the sequential circuit.

The algorithm consists of three main stages: vectorization, gathering, and multiplication. Each stage will be analyzed in detail in this Section, while the specific implementation of these stages, along with the measurement of ancillary qubits, is discussed in Section \ref{subsec:circuit}.

\subsection{Vectorization}

Consider two matrices $A,B$ of size $N \times N$. To obtain vectorization of these two operators we will use the following equation 
\begin{equation}\label{eq:vectorize}
     \vectorize{A B} = (B^T \otimes A) \vectorize{I_N},
\end{equation}
where $I_N$ is the identity matrix of size $N$ \cite{Phan_2009}.

Therefore, if we initialize quantum circuit in the state $\ket{\psi} = \frac{1}{\sqrt{N}}\vectorize{I_N}$ and apply the operator $(B^T \otimes A) $, the resulting state will be proportional to $\vectorize{A B}$.

\subsection{Gathering}\label{subsec:gather}

Using equation \eqref{eq:vectorize}, we can obtain the state, which is proportional to the tensor product of vectorized operators. In order to get the vectorization of the product of the considered operators, we introduce the following gathering operation $G$, which has the size of $N^2 \times N^4$
\begin{equation}
    G \left(\vectorize{A} \otimes \vectorize{B} \right) = \vectorize{BA}.
\end{equation}
In fact, we note that this operator can be written as 
\begin{equation}
    G = I_{N} \otimes \vectorize{I_{N}}^T \otimes I_{N}.
\end{equation}

Since $I_{N} \otimes \vectorize{I_{N}}^T \otimes I_{N}$ is not a square matrix, its direct use in a quantum algorithm is unclear. To address this, we define a square matrix $G_u$ of dimension $N^2 \times N^2$, where the first row is given by $\vectorize{I_{N}}^T$. There are various methods for constructing such matrix; for example, one approach to construct unitary $G_u$ is to generate a complete basis with the Gram-Schmidt process, starting with $\frac{1}{\sqrt{N}} \vectorize{I_{N}}$, and use these vectors as the rows of the matrix $G_u$. Alternatively, as discussed in Section \ref{subsec:implement}, we adopt a different approach where $G_u$, though not unitary itself, is represented as the sum of two unitary matrices.

For the case where $G_u$ is a matrix of dimension $N^2 \times N^2$, with the first row given by $\vectorize{I_{N}}^T$, we observe that applying this matrix to any tensor product of vectors places the desired state in the first and fourth registers (or equivalently, the first and last $n$ qubits). Specifically, we have

\begin{equation}\label{eq:Gu_final}
\begin{split}
    &\swapgate{2,4} \left[I_{N} \otimes G_u \otimes I_{N} \right] \left(\vectorize{A} \otimes \vectorize{B} \right) \\
     & =\vectorize{BA} \otimes \ket{0} \otimes \ket{0} + \sum_{\substack{k,l=0, \\ \text{except } \\ k=l=0}}^N \vec{r}_{k,l} \otimes \ket{k} \otimes \ket{l},
\end{split}
\end{equation}
where  $\swapgate{i,j}$ denotes a swap gate between the $n$-qubit registers $i$ and $j$, and $\vec{r}_{k,l}$ represents residual states that are not relevant to our current focus. Note that the states $\vectorize{BA}$ and $\vec{r}_{k,l}$ reside in the first two registers. The swap operator is introduced here for notational convenience but is not essential to the procedure itself.

\subsection{Multiplication} \label{subsec:multi}

The only part left to build our algorithm is the multiplication of a vector by a matrix. For a vector $\Vec{v}$ of size $N$, and a matrix $A$ of size $N \times N$, where $N=2^n$ we introduce the following operator $V$ of size $N \times N^2$

\begin{equation}
    V \vectorize{A^T} = A \Vec{v}.
\end{equation}

This operation can be implemented as 

\begin{equation}
    V = I_{N} \otimes \Vec{v}^T.
\end{equation}
Similarly to the gathering operator, we can extend the row $\vec{v}^T$ into a square matrix $V_u$ of size $N \times N$, whose first row is given by $\vec{v}^T$. Our method for constructing this matrix is provided in Section \ref{subsec:implement}. At this stage, we note that after applying $V_u$ to any vector, the desired state will be located in the first register (or the first $n$ qubits), i.e.
\begin{equation}\label{eq:Vu_final}
    \left[ I_{N} \otimes V_u \right] \vectorize{A^T} = (A \Vec{v}) \otimes \ket{0} + \sum_{k = 1}^N \ket{\vec{r}_{k}} \otimes \ket{k},
\end{equation}
where $\ket{\vec{r}_{k}}$ are some residual vectors, which we are not concerned with.

\section{Circuit for the algorithm} \label{subsec:circuit}

In this section, we first provide the circuit for the algorithm in terms of $G_u$ and $V_u$ and then we show how to realize them. Based on equations \eqref{eq:vectorize} and \eqref{eq:Vu_final} it is easy to see how one can create a circuit for parallelization of two gates. This circuit is presented in Figure \ref{fig:example_parseq_circ} (b). We formalize this in the following Proposition

\begin{proposition} [2 parallel gates]\label{prop:dql_2}
Consider a quantum circuit with two gates, $U_1$ and $U_2$, each of size $N \times N$, acting on an initial state $\ket{\Vec{v}}$. The circuit for parallel implementation can be expressed as:
\begin{equation}\label{eq:2_parallel_gates}
    V (U_1 \otimes U_2^T) \ket{\vectorize{I_N}} = \frac{1}{\sqrt{N}} U_1 U_2\ket{\Vec{v}} \otimes \ket{0} + \sum_{k = 1}^N \ket{\vec{r}_{k}} \otimes \ket{k},
\end{equation}
where $V = I_{N} \otimes V_u$, and $V_u$ is a unitary matrix whose first row is given by $\ket{\Vec{v}}^T$. The terms $\ket{\vec{r}_{k}}$ represent residual vectors, and $\ket{\vectorize{I_N}} = \frac{1}{\sqrt{N}} \vectorize{I_N}$, where $\vectorize{I_N}$ denotes the vectorization of the identity matrix of size $N$.
\end{proposition}

The gathering gate $G$ described in equation \eqref{eq:Gu_final}, can be applied to generalize the previous proposition to any $M > 2$. In the following, we assume $M$ is even; if not, we introduce $U_{M+1} = I$ to ensure even parity.

\subsection{Distributed quantum logic algorithm} \label{subsec:dql_alg}

The algorithm for constructing the quantum circuit comprises three primary stages: vectorization, gathering, and multiplication.  We also note that the use of swap operators is not necessary; we use them for the illustration purposes only.

\paragraph{Vectorization stage}
During this stage we construct $M/2$ vectorization circuits as defined by equation \eqref{eq:vectorize}. Specifically, the circuit is initialized in the state $\underbrace{\ket{\vectorize{I_N}} \otimes \dots \otimes \ket{\vectorize{I_N}}}_{M/2}$, and operators are applied, alternating transpositions, as $U_1~\otimes~U_2^T~\otimes~U_3~\otimes~U_4~\dots~\otimes~U_{M-1}~\otimes~U_M^T$. At the end of this stage, the quantum state becomes proportional to the tensor product of the vectorized operators:
\begin{equation*}
    \vectorize{(U_1 U_2)^T} \otimes \dots \otimes \vectorize{(U_{M-1} U_M)^T}.
\end{equation*}

\paragraph{Gathering stage}
Gathering operators $G~=~I_{N}~\otimes~G_u~\otimes~I_{N}$ are employed to combine the vectorized states. Consider a swap gate between the second and fourth output registers of the operator $G$ (recall that $G_u$ acts on two registers), producing the desired state in the first two output registers. Let us define the combination of $G$ and the swap gate as $\tilde{G}$. For simplicity, we will ignore the terms involving residual vectors $\ket{\vec{r}_{i}}$ in equation \eqref{eq:Gu_final}, focusing solely on the first term.

First let us consider the case $M = 2^\kappa$ (with $\kappa > 1$). After applying $\underbrace{\tilde{G} \otimes \dots \otimes \tilde{G}}_{M/4}$, the resulting state is proportional to:
\begin{equation*}
\vectorize{(\prod_{j=1}^4 U_j)^T} \otimes \ket{0 0} \otimes 
\dots \otimes \vectorize{(\prod_{j=M-3}^M U_j)^T} \otimes \ket{0 0}.
\end{equation*}
We then apply swap gates to bring the vectorized operators together, yielding the state:
\begin{equation*}
\vectorize{(\prod_{j=1}^4 U_j)^T} \otimes \vectorize{(\prod_{j=5}^8 U_j)^T} \otimes \ket{0 0 0 0} \otimes \dots.
\end{equation*}
This process is repeated until the state $\vectorize{(\prod_{j=1}^M U_j)^T} \otimes \underbrace{\ket{0} \dots \ket{0}}_{M-2}$ is obtained. The process requires $\left( \log_2(M) - 1\right)$ layers. In the first layer, there are $M/4 = 2^{\kappa-2}$ operators $\tilde{G}$; the second layer contains $M/(4 \times 2) = 2^{\kappa-3}$ operators, and so on, until the final layer, which consists of a single $\tilde{G}$ operator.

For any even $M$, the same method can be generalized by decomposing $M$ into sums of powers of two. The algorithm constructs circuits for each contributing $2^\kappa$, starting with the largest power. These circuits are then connected layer by layer, using swap and $\tilde{G}$ gates in a similar fashion as described for powers of two. The number of additional $\tilde{G}$ gates required corresponds to the number of $1$'s in the binary decomposition of $M$, minus $1$.

\paragraph{Multiplication stage}
The operator $V = I_{N} \otimes V_u$ is applied to the first two registers, completing the computation. The resulted circuit yields the final state, proportional to $\left( \prod_{j=1}^M U_j \right) \vec{v} \otimes \underbrace{\ket{0} \dots \ket{0}}_{M-1}$.

\begin{figure}[t]
\includegraphics[width=1\linewidth]{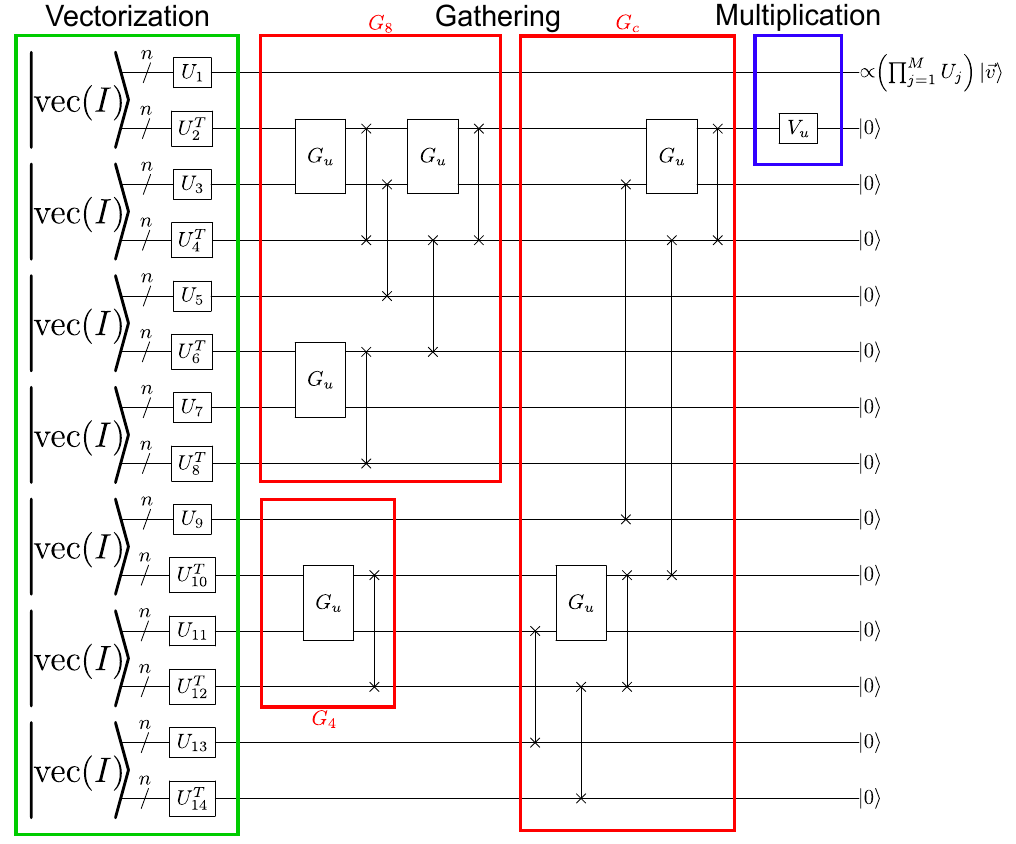}
\caption{\label{fig:par_circ_14} Circuit for parallel computation of $M = 14$ gates. The vectorization stage is highlighted by the green rectangle, while the gathering stage is represented by three red rectangles: $G_8$, $G_4$, and $G_c$. Here, $G_8$ and $G_4$ are gathering circuits for 8 and 4 gates, respectively, and $G_c$ is the gathering circuit that combines the remainders. The multiplication stage is indicated by the blue rectangle. The desired quantum state is illustrated on the right.}
\end{figure}

Figure \ref{fig:par_circ_14} provides a detailed illustration of the circuit configurations for $M~=~14$. This specific value is selected as it includes circuits for $2^3$ and $2^2$, highlighted with red rectangles and labeled as $G_8$ and $G_4$, respectively, as well as $G_c$, which represents the combining component of the gathering stage. This configuration demonstrates how each component integrates within the larger circuit.

The following proposition offers the scaling behavior for the circuit constructed using this algorithm.

\begin{proposition}[Distributed Qubit Logic (DQL)]\label{prop:dql_M}
    Consider $M > 1$ unitary operators $U_1, \ldots, U_M$, each of size $N \times N$ with $N = 2^n$, acting on an initial state $\ket{\vec{v}}$. To achieve a state proportional to $\left(U_1 \cdots U_M\right) \ket{\vec{v}}$ in the first register, one can construct a quantum circuit with a depth $d$ defined by:
    \begin{equation}\label{eq:dql_logic_scale}
    d = d_U + \left(\lceil \log_2(M) \rceil - 1\right) d_G + d_V,
    \end{equation}
    where $d_U$ represents the maximum depth of the operators $U_j$ for $j = 1, \ldots, M$, $d_G$ denotes the depth of the operator $G = I_N \otimes G_u \otimes I_N$, given by \eqref{eq:Gu_final}, and $d_V$ is the depth of the operator $V = I_{N} \otimes V_u$, given by \eqref{eq:Vu_final}. The number of $G_u$ gates required for the implementation of this algorithm is $M/2 - 1$, in addition to one $V_u$ gate. Furthermore, the total number of qubits required is $nM$.
\end{proposition}

\subsection{Implementation of $G_u$ and $V_u$} \label{subsec:implement}

In this section, we present an approach to implement the gates $G_u$ and $V_u$. We analyze their complexity in terms of one- and two-qubit gates. We begin with the implementation of the operator $G_u$. To construct $G_u$, we utilize the sum of Pauli strings (tensor products of Pauli matrices) that consist solely of Pauli matrices $X$ and $I$. Each such Pauli string can be expressed as $P_j = \bigotimes_{k=1}^{n} X^{(j)_k}$, where $(j)_k$ represents the $k$-th bit of $j$ in binary format, using reverse encoding (with the least significant bit on the left). The bitstrings are padded with zeros to the right as necessary, for $j = 0, \dots, 2^n-1$, with $X^0 = I$. Therefore, matrix $G_u$ with the first row given by $\vectorize{I_N}^T$ can be written as 
\begin{equation}\label{eq:def_Gu}
    G_u = \sum_{j=0}^{N-1} P_j \otimes P_j.
\end{equation}
This matrix is not unitary, so to implement on a quantum computer we will rewrite $G_u$ as
\begin{equation}
    G_u = \frac{N}{2} \left( I -e^{i \frac{\pi}{N} G_u} \right).
\end{equation}
This can be seen from the fact that $(P_j \otimes P_j) G_u = G_u$. Now, since all Pauli strings $P_j$ are commute, we can simultaneously diagonalize them, obtaining 
\begin{equation}
    G_u = \frac{N}{2} \left( I - (D^\dagger \otimes D^\dagger) e^{i \frac{\pi}{N} \hat{P}_j} (D \otimes D) \right),
\end{equation}
where $\hat{P}_j$ are diagonal matrices, namely Pauli string, which contains $Z$ and $I$ operators only, and $D$ is the diagonalization operator, namely $D = \underbrace{H \otimes \dots \otimes H}_{n} \equiv H^{\otimes n}$, where $H$ is the Hadamard gate. 

It is possible to further simplify this equation. To do this, one can use the following operator 
\begin{equation} \label{eq:S_gate}
    S \equiv \prod_{j=1}^n CX(j, j+n),
\end{equation}
where $CX(k, l)$ is denoting controlled $X$ gate with $k$ being the control qubit and l being the target qubit. We also note here that the depth of this operator is $1$.

Now it is possible to get the following result 
\begin{equation} \label{eq:Gu_implement}
    G_u = \frac{N}{2} \left( I - H^{\otimes 2n} S \left( I_{N} \otimes X_{n} C_0^nZ X_{n} \right) S H^{\otimes 2n} \right),
\end{equation}
where $C_0^nZ$ denoting multi controlled $Z$ gate that controls on a $0$, $X_{n} \equiv I^{\otimes n-1} \otimes X$, that is the $X$ gate which acts only on $n$-th qubit, and $I_{N}$ being an identity matrix of size $N = 2^n$.

\begin{figure}[t]
\includegraphics[width=1\linewidth]{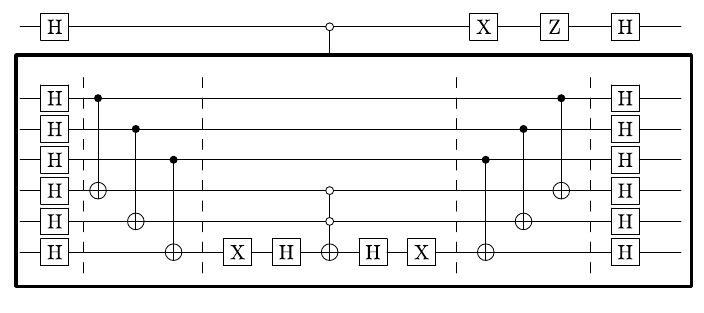}
\caption{\label{fig:G_u_circuit_n-3} Circuit for implementation $ \frac{1}{N} G_u = \frac{1}{2} \left(I -e^{i \frac{\pi}{N} G_u}\right)$, where $G_u$ is of size $N^2 \times N^2$, with $N=2^n$ and $n=3$. Here $C^2Z$ implemented as  $(I \otimes I \otimes H) C^2X (I \otimes I \otimes H)$.}
\end{figure}

Finally, we can apply a technique called the linear combination of unitaries (LCU) \cite{childs2012hamiltonian}. The finial complexity of implementation of $G_u$ is: $4n$ controlled Hadamard gates, $2n$ $CCX$ gates, $2$ $CX$ gates, $2$ Hadamard gates, and one $C_0^{n+1}Z$ gate. Controlled Hadamard gate can be implemented as $CH = (I \otimes R_y(\frac{\pi}{4})) CZ (I \otimes R_y(-\frac{\pi}{4}))$, $CCX$ can be implemented with $6$ $CX$ gates and $C_0^{n+1}Z$ can be implemented recursively with $O(n^2)$ gates and an auxiliary gubit \cite{barenco1995elementary,iten2016quantum}. Taking all this into account, the total complexity of implementation of $G_u$ strongly depends on the implementation of $C_0^{n+1}Z$ gate. In fact this implementation may be linear but will require a lot of ancillary qubits \cite{barenco1995elementary,iten2016quantum}. The circuit for implementation of $G_u$ up to a constant factor for the case of $n=3$ is presented in Figure \ref{fig:G_u_circuit_n-3}. We modified the original LCU method with X and Z gates in the first register, in order to obtain desired result with ancillary qubit being in state $\ket{0}$, that is the output state of the circuit from Figure \ref{fig:G_u_circuit_n-3} is 
$\frac{1}{2} \left(\ket{0}\left(I -e^{i \frac{\pi}{N} G_u}\right) \ket{\psi} + \ket{1}\left(I + e^{i \frac{\pi}{N} G_u}\right) \ket{\psi}\right)$, where $\ket{0}\ket{\psi}$ is initial state.

Next, we consider the operator $V_u$. Similarly to $G_u$ the operator which contain $\Vec{v}^T$ as a first row can be written as
\begin{equation}
    V_u = \sum_{j=0}^{N-1} v_j P_j,
\end{equation}
where $v_j$ are the components of $\Vec{v}$. Now it is possible to diagonolize it as 
\begin{equation}
    V_u  = D^\dagger \Lambda_V D,
\end{equation}
where the diagonalization operator is the same as for $G_u$, that is $D = H^{\otimes n}$. Next, we apply $\arccos$ function to the elements of $\Lambda_V / \lambda_{\max}$, where $\lambda_{\max} \equiv \sum_{j=0}^{N-1} \abs{v_j}$ is needed for normalization.
The resulted matrix $\Lambda_{\arccos}$ will have elements $\arccos(\lambda_j/ \lambda_{\max})$ on the diagonal. With the help of this matrix we can reconstruct $\Lambda_V$ as a sum of $2$ unitary matrices 
\begin{equation}
    \Lambda_V = \frac{1}{2} \left( e^{i \Lambda_{\arccos}} + e^{-i \Lambda_{\arccos}}\right),
\end{equation}
that is 
\begin{equation}
    \frac{V_u}{\lambda_{\max}} =   H^{\otimes n} \frac{1}{2} \left( e^{i \Lambda_{\arccos}} + e^{-i \Lambda_{\arccos}}\right) H^{\otimes n}.
\end{equation}
In order to implement each term in this formula, one can use results from \cite{Welsh} where it is shown that the gate count for a circuit which implements the exponent of a diagonal can be reduced to $O(N')$ gates with $N' < 2^n$, without ancillary qubits. Therefore, the total gate count for the implementation of $V_u$ is $O(N)$. 

We formulate the obtained results in the Proposition \ref{prop:GV_implement}.

\begin{proposition}[Implementation of $G_u$ and $V_u$] \label{prop:GV_implement}
Gate $G_u$ of size $N^2 \times N^2$, with $N = 2^n$ such that its first row is proportional to $\vectorize{I_N}^T$ can be implemented with $O(n^2)$ gates and 2 ancillary qubits.
Gate $V_u$ of size $N \times N$, with $N = 2^n$ such that it has $\ket{\Vec{v}}^T$ as a first row can be implemented with $O(N)$ gates and 1 ancillary qubit.
\end{proposition}

We note that in general, the implementation of the initial condition $\ket{\Vec{v}}$, where $\Vec{v}$ is vector of size $N=2^n$ has the complexity of $O(N)$ \cite{shende2005synthesis, mahmud2021optimizing}. In other hand, the initial condition $\ket{\vectorize{I_N}}$ can be implemented as 
\begin{equation}
    \ket{\vectorize{I_N}} = S \left( H^{\otimes n} \ket{0}^{\otimes n} \otimes \ket{0}^{\otimes n} \right),
\end{equation}
where $S$ gate is given by \eqref{eq:S_gate}. This implementation contains only $n$ controlled $X$ gates and $n$ Hadamard gates. The depth of such implementation is $2$. Therefore, if we take into account that the cost of implementation of $\ket{\Vec{v}}$ is the same as for implementation of $V_u$ we can consider only the cost of implementation of the circuit itself. We formulate this result in the following Corollary \ref{cor:effiecient}. 

\begin{corollary}[Efficient parallelization] \label{cor:effiecient}
Any sequence of gates on $n$ qubits of depth $O(M n^2)$ can be represented as a circuit of depth $O(\log_2(M) n^2)$ acting on $O(Mn)$ qubits.
\end{corollary}

This result enables the construction of a circuit with logarithmic depth relative to the system size $N$, indicating that any problem expressible as a sequence of gates can be solved in logarithmic time, assuming a sufficient number of qubits is available. A challenging aspect of this algorithm, however, lies in the probability of obtaining the desired state. We address this issue below.

\subsection{Probability of getting the state} \label{subsec:success}

First, we recall the matrix $G_u$ defined in equation \eqref{eq:def_Gu}, $G_u = \sum_{j=0}^{N-1} P_j \otimes P_j$, where $P_j = \bigotimes_{k=1}^{n} X^{(j)_k}$. Note that $G_u$ can be expressed as

\begin{equation}\label{eq:Gu_structure}
G_u = 
    \begin{pmatrix}
        P_0 & P_1 & \dots & P_{N-1} \\
        P_1 P_0 & P_1 P_1 & \dots & P_1 P_{N-1} \\
        \vdots & \vdots & \dots & \vdots\\
        P_{N-2} P_0 & P_{N-2} P_1 & \dots & P_{N-2} P_{N-1} \\
        P_{N-1} P_0 & P_{N-1} P_1 & \dots & P_{N-1} P_{N-1} \\
    \end{pmatrix},
\end{equation}

and furthermore,

\begin{equation}\label{eq:Gu_rows}
    \begin{pmatrix}
        P_j P_0 & P_j P_1 & \dots & P_j P_{N-1}
    \end{pmatrix}
    =
    \begin{pmatrix}
        \vectorize{P_j P_0}^T \\
        \vectorize{P_j P_1}^T \\
        \vdots \\
        \vectorize{P_j P_{N-1}}^T
    \end{pmatrix}.
\end{equation}

That is each row of matrix $G_u$ can be written as $\vectorize{P_j}$. Recalling equation \eqref{eq:Gu_final},

\begin{equation*}
\begin{split}
    &\swapgate{2,4} \left[I_{N} \otimes G_u \otimes I_{N} \right] \left(\vectorize{A} \otimes \vectorize{B} \right) \\
     & =\vectorize{BA} \otimes \ket{0} \otimes \ket{0} + \sum_{\substack{k,l=0, \\ \text{except } \\ k=l=0}}^{N-1} \vec{r}_{k,l} \otimes \ket{k} \otimes \ket{l},
\end{split}
\end{equation*}

we observe, using equations \eqref{eq:Gu_structure} and \eqref{eq:Gu_rows}, that

\begin{equation}
    \vec{r}_{k,l} = \left[I_{N} \otimes \vectorize{P_k P_l}^T \otimes I_{N} \right] \left(\vectorize{A} \otimes \vectorize{B} \right).
\end{equation}

Applying another property,

\begin{equation*}
\begin{split}
    I \otimes \vectorize{P_j}^T \otimes I &= (I \otimes \vectorize{I}^T \otimes I) (I \otimes P_j \otimes I \otimes I) \\
    &= (I \otimes \vectorize{I}^T \otimes I) (I \otimes I \otimes P_j \otimes I),
\end{split}
\end{equation*}

we can rewrite

\begin{equation*}
\begin{split}
    (I \otimes \vectorize{P_k P_l}^T \otimes I) (\vectorize{A} \otimes \vectorize{B}) = \vectorize{B P_k P_l A}.
\end{split}
\end{equation*}

Thus, equation \eqref{eq:Gu_final} for the specific $G_u$ under consideration can be expressed as

\begin{equation*}
\begin{split}
    &\swapgate{2,4} \left[I_{N} \otimes G_u \otimes I_{N} \right] \left(\vectorize{A} \otimes \vectorize{B} \right) \\
     & =\sum_{k,l=0}^{N-1} \vectorize{B P_{k} P_{l} A} \otimes \ket{k} \otimes \ket{l}.
\end{split}
\end{equation*}

Given the measurement outcomes (i.e., values $k$ and $l$), we can identify the specific vector $\vectorize{B P_k P_l A}$ applied. Consequently, if the operators $U$ commute with the Pauli strings $P_j$, it becomes possible to obtain the exact result from any measurement by applying the corresponding Pauli string.

For the matrix $V_u$, a similar scenario holds, allowing Equation \eqref{eq:Vu_final} to be expressed as
\begin{equation}
    \left[ I_{N} \otimes V_u \right] \vectorize{A^T} = \sum_{j = 0}^{N-1} (A P_{j} \Vec{v}) \otimes \ket{j}.
\end{equation}

Now, let us analyze the impact of the Linear combination of unitaries on the implementation of $G_u$. For a given state $\ket{\vec{w}}$, the norms are given by
\begin{equation}
\begin{split}
    \left\|\frac{1}{2}\left( I - e^{i \frac{\pi}{N} G_u}\right) \ket{\vec{w}} \right\|^2 &= \frac{W^2}{N}, \\
    \left\|\frac{1}{2}\left( I + e^{i \frac{\pi}{N} G_u}\right) \ket{\vec{w}} \right\|^2 &= 1 - \frac{W^2}{N},
\end{split}
\end{equation}
where $\ket{\vec{w}} = (w_0, \dots, w_{N-1})^T$ and $W = \sum_{j=0}^{N-1} w_j$. This result suggests that the probability of obtaining the correct LCU outcome depends on the specific vector $\ket{\vec{w}}$. For example, when $\ket{\vec{w}} = \frac{1}{\sqrt{N}}(1, \dots, 1)^T$, $W$ reaches its maximum of $\sqrt{N}$.

The effect of LCU on the implementation of $V_u$ is more challenging to evaluate precisely. While we can express it as follows, this expression does not easily reveal insights about performance:
\begin{equation*}
\begin{split}
    &\norm{H^{\otimes n} \frac{1}{2} \left( e^{i \Lambda_{\arccos}} + e^{-i \Lambda_{\arccos}}\right) H^{\otimes n} \ket{\vec{w}}}^2 \\
    &= \frac{1}{\lambda_{\max}^2} \left( \sum_{j=0}^{N-1} (P_j \vec{v},\vec{w})^2\right).
\end{split}
\end{equation*}

This expression highlights that the behavior depends on the overlap between $P_j \vec{v}$ and $\vec{w}$ for each $j$. However, interpreting the impact on the success probability directly from this form remains difficult without specific properties of $\vec{v}$ and $\vec{w}$. Notably, in the algorithm, only a single $V_u$ gate is employed, requiring just one ancilla qubit for the LCU implementation of this gate.

\section{Conclusion}
In this work, we have introduced a universal method for reducing the depth of quantum circuits by leveraging additional qubits, referred to as the Distributed Quantum Logic (DQL) algorithm. This method distributes quantum gates across different registers, enabling parallel execution of a sequence of $M$ operators $U$ through the use of specialized gates, denoted as $G$ and $V$. The number of $G$ gates required is $(M/2 - 1)$ in addition to one $V$ gate, resulting in an overall gate count of $\frac{3}{2}M$ for $G$, $U$, and $V$ gates. The main result is a reduction in circuit depth to $\left( \lceil \log_2(M) \rceil + 1 \right)$ in terms of these gates. This is demonstrated for $M=2$ in Proposition \ref{prop:dql_2} and generalized in Proposition \ref{prop:dql_M}.

Additionally, we provide a method for implementing the $G$ and $V$ gates, formalized in Proposition \ref{prop:GV_implement}. The circuit required for the implementation of $G$ 
consists of $O(n^2)$ gates and $2$ additional qubits, while for $V$ the circuit consists of $O(N)$ gates, where $N=2^n$ and uses $1$ additional qubit. As a result, for any circuit on $n$ qubits with depth $O(M n^2)$, where $M > 1$, our algorithm reduces the circuit depth to $O(\log_2(M) n^2)$, as shown in Corollary \ref{cor:effiecient}. A challenging aspect of this implementation lies in the probability of obtaining the desired state. We showed, that using proposed implementation of $G$, we know which operator was applied based on measurement of ancilla qubits, namely where in the resulted sequence Pauli matrices were applied. We believe that other techniques exist or may appear to implement matrices $G_u$ without having this issue.

During the development of this algorithm, we adopted a distinct perspective on quantum circuits, where the initial state is treated as a gate and operators are represented as vectors. This approach offers potential advantages for a range of quantum computing applications. Specifically, the DQL algorithm could be beneficial for Hamiltonian simulation, where the Trotter formula often leads to long circuits. For instance, in our previous work \cite{tridiagonal}, we achieved a scalability of $O(N^{2}n^{5/2})$ for solving the wave equation on $n$ qubits, which can be improved to $O(\log_2(N^2 n^{1/2})n^2)$ = $O((2n+1/2 \log_2(n))n^2) = O(n^3)$ using the DQL algorithm with $O(N^2 n^{3/2})$ additional qubits. Furthermore, this method may have applications in variational quantum algorithms, where it has been shown that in most of the circuits with $n$ qubits, only the final $O(\log(n))$ layers significantly influence the expectation values of observables \cite{mele2024noise}.

The PYTHON code with the numerical experiments is available on \href{https://github.com/barseniev/General-Matrix-Decomposition}{GitHub} \cite{github_page}.

\bibliography{apssamp}% Produces the bibliography via BibTeX.

\end{document}